\documentclass[aps,prd,preprint,tightenlines,floatfix]{revtex4}
\usepackage{graphicx}

\begin{document}
\preprint{BI-TP 2008/13}
\date{Nov 21, 2008}
\title{The effect of early dark matter halos on reionization}

\author{Aravind Natarajan}
\email{anatarajan@physik.uni-bielefeld.de}
\affiliation{Fakult\"{a}t f\"{u}r Physik, Universit\"{a}t Bielefeld, Universit\"{a}tsstra\ss e 25, Bielefeld 33615, Germany}

\author{Dominik J. Schwarz}
\email{dschwarz@physik.uni-bielefeld.de}
\affiliation{Fakult\"{a}t f\"{u}r Physik, Universit\"{a}t Bielefeld, Universit\"{a}tsstra\ss e 25, Bielefeld 33615, Germany}

\begin{abstract} 
The annihilation of dark matter particles releases energy, ionizing some of the gas in the Universe. We investigate the effect of dark matter halos on reionization. We show that the effect depends on the assumed density profile, the particle mass, and the assumed minimum halo mass. For Navarro-Frenk-White halos and typical WIMPs, we find the effect to be quite small. However, light dark matter candidates in the MeV range can contribute significantly to reionization and can make an important contribution to the measured optical depth. This effect may be used to constrain light dark matter models. We also study the effect of varying the halo density profile on reionization.
\end{abstract}

\maketitle

  \section{Introduction.}
The absence of significant Ly$\alpha$ absorption (the Gunn Peterson test\cite{gp}) in the spectrum of many quasars implies that the Universe is highly ionized up to a redshift $\approx 6$. Observations of the cosmic microwave background (CMB) by the Wilkinson Microwave Anisotropy Probe (WMAP)\cite{wmap5} suggest that the Universe was reionized at a redshift $\approx 11$, assuming full ionization at lower redshifts. Primordial stars and quasars are commonly believed to have played a dominant role in the reionization of the Universe.  In this article, we investigate another possibility, namely whether radiation from the earliest dark matter halos could have contributed significantly to reionization.

If dark matter particles in halos annihilate producing standard model particles, some of the released energy is absorbed by the gas, resulting in ionization. The effect of particle annihilation on the ionization of gas by a uniform distribution of dark matter was studied by \cite{dm1,dm2}. These authors however, concluded that WIMP dark matter is unlikely to have a significant effect on ionization.  The effect of dark matter clumping was taken into account by \cite{dm3}, who modified the dark matter distribution by including a ``boost factor'', and found that WIMP annihilation could be relevant to reionization, and could account for part of the observed optical depth, particularly for light dark matter candidates. In this article, we extend the analysis of \cite{dm3}, by considering dark matter halos with a generalized Navarro-Frenk-White (NFW) profile. The number density of halos is determined by the Press-Schechter formula. We provide a detailed computation of the ionized gas fraction, and calculate the resulting optical depth, for different halo and particle parameters. We then show that our results may be used to constrain light dark matter candidates.

\subsection{Luminosity of halos.}

We fit each dark matter halo with a generalized NFW profile\cite{nfw}:
\begin{equation}
\rho(r) = \frac{\rho_{\rm s}}   { (r/r_{\rm s})^ \alpha \left[1 + r/r_{\rm s}\right ]^\beta}
\label{nfw}
\end{equation}
$\alpha=1, \beta = 2$ corresponds to the well known NFW profile\cite{nfw}. $\rho(r)$ is the dark matter density at $r$, and $\rho_{\rm s}$ and $r_{\rm s}$ are constants. Let $r_{\rm 200}$ denote the radius at which the mean matter density $\bar\rho$ equals 200 times the cosmological average at the formation redshift $z_{\rm F}$, i.e., 
\begin{equation}
\bar\rho(z_{\rm F}) = 200 \, \rho_{\rm c}  \, \Omega_{\rm m} \, (1+z_{\rm F})^3
\label{rho_bar}
\end{equation}
where $\rho_{\rm c} = 3 H^2_{\rm 0} / 8 \pi G$ is the critical density, $H_{\rm 0}$ is the Hubble parameter today and $\Omega_{\rm m}$ is the matter fraction. $z_{\rm F}$ is the redshift at which a 1-$\sigma$ fluctuation containing mass $M_{\rm min}$ enters the non-linear regime\cite{ghs1}. The minimum mass $M_{\rm min}$ is determined by the free streaming scale, which depends on the coupling of dark matter particles with standard model particles\cite{ghs1}. The mass in dark matter enclosed within $r_{\rm 200}$
\begin{eqnarray}
M_{\rm dm}(r_{\rm 200}) &=& f_{\rm dm} \, M \nonumber \\
  &=&  \frac{4 \pi}{3} r^3_{\rm 200} \, f_{\rm dm} \; \bar\rho(z_{\rm F})
\end{eqnarray}
$M = M(r_{\rm 200})$ is the halo mass. The concentration parameter $c_{\rm 200} = r_{\rm 200}/r_{\rm s}$ and $f_{\rm dm}$ is the fraction of mass in dark matter which we set equal to $\Omega_{\rm dm}/\Omega_{\rm m} = 0.8287$\cite{wmap5}. Note that we have defined $c_{\rm 200}$ at the time of halo formation $z_{\rm F}$. The luminosity of the halo per unit photon energy is then given by the expression:
\begin{eqnarray}
\frac{dL}{dx} &=& L_{\rm 0} (M, c_{\rm 200}) \times \frac{dN_\gamma}{dx} \, x \nonumber \\
L_{\rm 0} &=& \frac{\langle \sigma_{\rm a} v \rangle}{2\,m_{\rm dm}}  \, \int dr \, 4 \pi r^2 \, \rho^2(r)
\end{eqnarray}
where $x = E_\gamma / m_{\rm dm}$ and $m_{\rm dm}$ is the particle mass, measured in units of energy. $\langle \sigma_{\rm a} v \rangle$ is the annihilation cross section of the WIMPs times the relative velocity, averaged over the velocity distribution. We set $\langle \sigma_{\rm a} v \rangle = 3 \times 10^{-26}$ cm$^3$/s, in order to obtain $\Omega_{\rm dm} h^2\approx 0.1$ today\cite{jkg}. We will also assume that $\langle \sigma_{\rm a} v \rangle$ is independent of $v$.  $dN_\gamma/dE_\gamma = m^{-1}_{\rm dm} \, dN_\gamma/dx$ is the number of photons released per annihilation per photon energy. Here, we restrict ourselves to tree level processes. We use the phenomenological result of \cite{pheno} for neutralinos, and express $dN_{\rm \gamma}/dx = a\,e^{-bx} / x^{1.5}$, where $a$ and $b$ are constants for a particular annihilation channel. Averaging over the channels considered in \cite{pheno}, we find $a = 0.9, b = 9.56$.  The factor of 2 in the denominator accounts for the fact that two particles disappear per annihilation. Using Eq. \ref{nfw}, we find
\begin{eqnarray}
\int dr \, 4 \pi r^2 \, \rho^2(r) &=& \frac{M \, \bar\rho}{3} \left( \frac{\Omega_{\rm dm}}{\Omega_{\rm m}} \right )^2 \, f(c_{\rm 200}) \nonumber \\
f(c_{\rm 200}) &=& \frac{c^3_{\rm 200} \int_{\epsilon} ^{c_{\rm 200}} dx \, x^{2-2\alpha} \; (1+x)^{-2\beta}} { \left[ \int_{0}^{c_{\rm 200}} dx \, x^{2-\alpha} \; (1+x)^{-\beta} \right ]^2 }
\end{eqnarray}
where $\epsilon$ is a dimensionless cutoff scale, required to make the luminosity finite for $\alpha > 1.5$. For the NFW profile, we have $\alpha = 1, \beta = 2, \epsilon = 0$, and
\begin{equation}
f(c_{\rm 200}) = \frac{ c^3_{\rm 200} \, \left[ 1 - (1+c_{\rm 200})^{-3}  \right ] }{ 3 \, \left[ \log (1+c_{\rm 200}) - c_{\rm 200}(1+c_{\rm 200})^{-1} \right ]^2 }
\end{equation}

\section{Ionization of gas.} 

The probability of ionization per unit volume at location $s$ is given by
\begin{equation}
p(s) = \frac{ n_{\rm b} \, \left[ 1-x_{\rm ion}(s) \right ]  \left[ 1+z(s) \right] ^3  \, \sigma}{ 4 \pi s^2 }
\end{equation}
$s$ is the distance travelled by a light ray from redshift infinity to redshift $z$ (in the matter dominated era, $z \gg 1$)
\begin{equation}
s = \frac{2c}{H_{\rm 0}\sqrt{\Omega_{\rm m}}} \, \frac{1}{\sqrt{1+z}}
\label{sz}
\end{equation}
$\sigma$ is the scattering cross section of photons with gas atoms. $x_{\rm ion}$ is the ionized fraction and $n_{\rm b} \, [1-x_{\rm ion}]$ is the comoving number density of bound atoms. For a mixture of $76\%$ H and $24\%$ He, and assuming singly ionized He (the contribution of doubly ionized He is smaller, and is hence neglected here), we have
\begin{eqnarray}
n_{\rm b} &=& \frac{0.82 \rho_{\rm c} \, \Omega_{\rm b}}{m_{\rm p}} \nonumber \\  
\frac{di}{dE_\gamma} &\approx& \eta \, p(s) \, \mu \, \frac{dL}{dE_\gamma} 
\label{i_ch}
\end{eqnarray}
where
\begin{equation}
\mu = \left[ \frac{0.76}{0.82} \; \frac{1}{13.6 \,\textrm{eV}} + \frac{0.06}{0.82} \; \frac{1}{24.6 \,\textrm{eV}} \right ]  \approx 0.07 \; \rm{eV}^{-1}
\label{mu}
\end{equation}
$i(s)$ is the number of ionizations per unit time per unit volume, at $s$. For simplicity, we have accounted for the different ionization potentials of H and He using $\mu$ defined in Eq. \ref{mu}, rather than by modifying the cross section. In any case, $i(s)$ in Eq. \ref{i_ch} is only an approximation, since we do not consider in detail, the scattering processes involved. $\eta$ is the fraction of the energy absorbed by the gas, resulting in ionization. We assume $\eta = 0.3$, in accordance with \cite{shull}. We have assumed an ionization potential of 13.6 eV for Hydrogen and 24.6 eV for singly ionized Helium\cite{bark}. $\Omega_{\rm b}$ is the baryon fraction and $m_{\rm p}$ is the proton mass.  The cross section due to scattering of gas atoms by photons is the energy dependent Klein-Nishina cross section $\sigma(E_\gamma) = f_\sigma(E_\gamma) \, \sigma_{\rm T}$ \cite{ps} (For photon wavelengths much smaller than the Bohr radius, we may neglect details of the bound structure while computing the scattering cross section. We therefore consider only electron-photon scattering.):
\begin{equation}
f_\sigma(y) = \frac{3}{8}  \left[ \frac{2(1+y)}{(1+2y)^2} + \frac{\log(1 + 2y)}{y} - \frac{2}{1+2y} + \frac{2(1+y)^2}{y^2(1+2y)} - \frac{2(1+y) \log(1+2y)}{y^3} + \frac{2}{y^2} \right ]
\end{equation}
where $y = E_\gamma / m_{\rm e}, \sigma_{\rm T}$ is the Thomson cross section, and $m_{\rm e}$ is the electron mass.

The fraction of the number of photons available at location $s$, having been emitted at $s^{'}$ 
\begin{equation}
\kappa(s,s') = \exp \left[ - \sigma \, n_{\rm b}  \,  \int_{s'}^s \, ds'' \left[1 + z(s'') \right]^3 \right ]
\label{eta}
\end{equation}
In Eq. \ref{eta}, we have included both bound and ionized fractions in $n_{\rm b}$. Let $dn/dM$ be the comoving number density of dark matter halos per unit halo mass, at a redshift $z$, given by the Press-Schechter formula\cite{press}:
\begin{equation}
\frac{dn}{dM} = \sqrt{\frac{2}{\pi}} \, \frac{\rho_{\rm m}}{M} \, \frac{\delta_{\rm c} \, (1+z)}{\sigma^2_{\rm h}} \, \frac{d\sigma_{\rm h}}{dM} \, \exp \left[ -\frac{\delta^2_{\rm c} (1+z)^2}{2 \sigma^2_{\rm h}} \right ] 
\end{equation}
$\rho_{\rm m}$ is the comoving matter density. $\sigma^2_{\rm h}$ is the variance of the density field smoothed over scale $R$\cite{cs} (corresponding to a mass $M$)
\begin{eqnarray}
\sigma^2_{\rm h}(R) &=& \int \frac{dk}{k} \, \frac{k^3 P(k)}{2 \pi^2} \, \left | W(kR) \right |^2  \nonumber \\
W(kR) &=& \frac{3}{(kR)^3} \, \left[ \sin kR - (kR) \cos kR \right ]
\end{eqnarray}
$P(k)$ is the matter power spectrum which takes the form given by Eisenstein and Hu\cite{eh}, normalized to $\sigma_{\rm 8} = 0.8$. Following\cite{ba}, we set $\delta_{\rm c} = 1.28$. We can now write an expression for the number of ionizations per unit volume per unit time, at redshift $z$:
\begin{equation}
I(z) = A \, \mu \, \eta \, \left[1 - x_{\rm ion} \right ] (1+z)^5 \,  \int_{z_{\rm F}}^z \, -dz' \, (1+z')^{-1/2} \int_0^1 dx \frac{a e^{-bx}}{\sqrt{x}} f_\sigma \, S \int_{M_{\rm min}}^\infty \, dM \, \frac{dn}{dM} \, L_{\rm 0}
\label{I}
\end{equation}
In Eq. \ref{I}, we have included all halos with masses above the cutoff mass $M_{\rm min}$. We then integrated over photon energies, and finally over redshift, from the formation redshift $z_{\rm F}(M_{\rm min})$ to $z$. The factors of $(1+z)$ occur because we made a change from co-ordinate space to redshift space, and also account for the redshifting of a photon as it propagates from $z'$ to $z$. $A$ is the dimensionless quantity 
\begin{equation}
A = \frac{c \, \sigma_{\rm T}  n_b}{H_{\rm 0} \sqrt{\Omega_{\rm m}}} \approx 3.5 \times 10^{-3}
\end{equation}
and $S(z',z; E_\gamma)$ is given by the expression
\begin{equation}
S = \exp \left[ - \frac{2 A f_\sigma}{5} \left\{ (1+z')^{5/2} - (1+z)^{5/2} \right \} \right ]
\end{equation}

Let us now consider recombination into bound atoms. The recombination rate per unit volume 
\begin{equation}
R(z) =  n^2_{\rm b}\, x^2_{\rm ion} (1+z)^6 \left[ \frac{0.76}{0.82} \, \alpha_{\rm H} + \frac{0.06}{0.82} \, \alpha_{\rm He} \right ]
\label{rec}
\end{equation}
$\alpha_{\rm H}$ and $\alpha_{\rm He}$ are given by\cite{abel}
\begin{eqnarray}
\alpha_{\rm H} &\approx& 3.746 \times 10^{-13} (T/\rm{eV})^{-0.724} \;\; \textrm{cm}^3 \, \textrm{s}^{-1}\nonumber \\
\alpha_{\rm He} &\approx& 3.925 \times 10^{-13} (T/\rm{eV})^{-0.6353} \;\; \textrm{cm}^3 \, \textrm{s}^{-1} 
\end{eqnarray}
$T$ is the gas temperature $\approx 8 \times 10^{-4} [(1+z)/61]^2$ eV. Using Eq. \ref{I} and Eq. \ref{rec}, we solve for $x_{\rm ion}(z)$ 
\begin{eqnarray}
I(z) - R(z) &=& n_{\rm b} (1+z)^3 \frac{dx_{\rm ion}}{dt} \nonumber \\
     &=& - n_{\rm b} H_{\rm 0} \sqrt{\Omega_{\rm m}}  \, \frac{dx_{\rm ion}}{dz} \, (1+z)^{11/2}
     \label{ion}
 \end{eqnarray}
where we have used the relation 
\begin{equation}
- H_{\rm 0} \, \sqrt{\Omega_{\rm m}} \; dt = \frac{dz}{ (1+z) \sqrt{ (1+z)^3 + \frac{\Omega_{\rm \Lambda}}{\Omega_{\rm m}} } }
\label{zt}
\end{equation}
with the approximation $\Omega_{\rm \Lambda} \ll \Omega_{\rm m} (1+z)^3$, valid at high redshifts. We neglect gas clumping in this work. The effect of gas clumping  would be to increase the recombination rate. We however expect this effect to be small at redshifts $z > 10$.

\section{Optical depth.} 

As mentioned earlier, most of the Hydrogen in the Universe is ionized up to redshift $z=6$. We assume that Helium is singly ionized at $z=6$ and doubly ionized\cite{helium} at $z = 3$. We then have the optical depth
\begin{equation}
\tau(z<6) = n_{\rm b} \, \sigma_{\rm T} \left[ -\frac{0.88}{0.82} \, \int_0^3  dz \,  \frac{dt}{dz} (1+z)^3 - \int_3^6  dz \,  \frac{dt}{dz} (1+z)^3   \right ]
\end{equation}
which yields $\tau(z<6) = 0.04$ where we have used Eq. \ref{zt} and the 5 year mean values of the WMAP experiment\cite{wmap5} $H_{\rm 0} = 71.9, \Omega_{\rm \Lambda} = 0.742, \Omega_{\rm m} = 0.258, \Omega_{\rm b} = 0.04422$. From the mean value of the WMAP measured optical depth\cite{wmap5} $\tau = 0.087\pm0.017$, we note that an excess $\delta\tau \approx 0.047$ needs to be provided by sources at redshift $z > 6$. We explore the dark matter models that can contribute this value $\delta\tau$ to the optical depth. In these models, $\delta\tau$ is provided solely by dark matter halos. 

\section{Results.}

We now present our results for $x_{\rm ion}$ and $\delta\tau$. We set $\langle \sigma_{\rm a} v \rangle = 3 \times 10^{-26}$ cm$^3$ s$^{-1}$ and the density profile power law exponent $\beta = 2$ throughout. Our results thus depend upon $\alpha$, the cutoff parameter $\epsilon$, the halo concentration parameter $c_{\rm 200}$, the particle mass $m_{\rm dm}$, and the minimum halo mass $M_{\rm min}$. For simplicity, we treat $c_{\rm 200}$ as a free parameter, and vary it independently of the halo mass.

\subsection{A generic WIMP.}

\begin{figure}[!ht]
\begin{center}
\scalebox{0.68}{\includegraphics{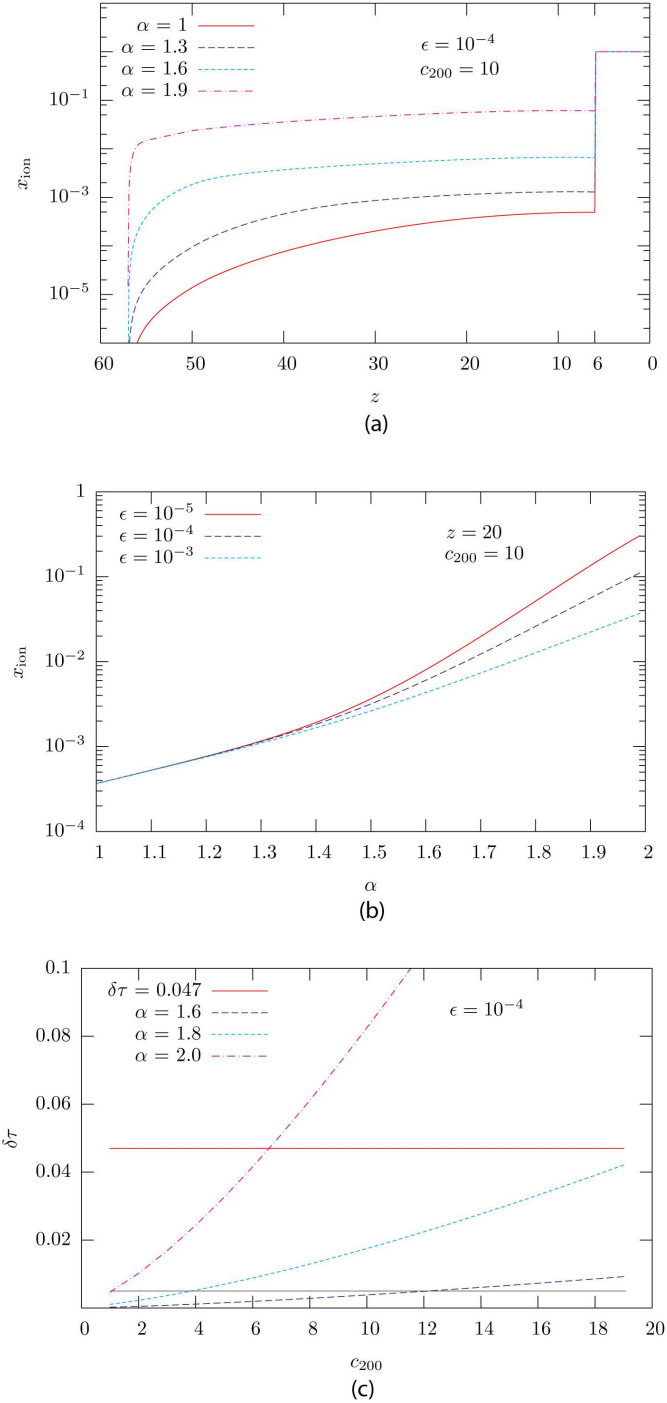}}
\end{center}
\caption{  (color online) Effect of neutralino dark matter halos on reionization and optical depth. (a)  shows $x_{\rm ion}$ for different values of the power law exponent $\alpha$. $z_{\rm F} \approx 57$ for $M_{\rm min} = 10^{-6} M_\odot$. $x_{\rm ion}$ is set to 1 for $z < 6$. (b) shows the variation of $x_{\rm ion}$ with the cutoff parameter $\epsilon$. The dependance on $\epsilon$ is very slight for $\alpha < 1.5$, but becomes important for larger values of $\alpha$. (c) shows the contribution of dark matter halos to the optical depth $\delta\tau$, for different values of $c_{\rm 200}$. The solid red line indicates the maximum value of $\delta\tau$ consistent with the WMAP observation. The solid black line indicates the expected sensitivity of the Planck experiment.\label{fig1} }
\end{figure}

The earliest WIMPy halos are expected to form at a redshift $z \sim 60$, with masses $M \sim 10^{-6} M_\odot$ \cite{ghs1,ghs2}. We now calculate the ionized fraction $x_{\rm ion}$ using Eq. \ref{ion}, and assuming a particle mass of $m_{\rm dm} = 100$ GeV, and $M_{\rm min} = 10^{-6} M_\odot$. $x_{\rm ion}$ depends on the power law exponent $\alpha$, cutoff $\epsilon$, and concentration parameter $c_{\rm 200}$. Fig \ref{fig1}(a) shows $x_{\rm ion}$ as a function of redshift, for 4 different values of $\alpha$, with $c_{\rm 200} = 10$, and $\epsilon = 10^{-4}$. $x_{\rm ion}$ was set equal to 1 for $z > 6$, in accordance with quasar studies. The NFW profile results in very little ionization, the peak ionization for $z < 6$ being only $5 \times 10^{-4}$. The ionized fraction increases with increase in $\alpha$. For the case $\alpha = 1.9$, we find a peak ionized fraction of 6\%. Fig. \ref{fig1}(b) shows $x_{\rm ion}$ at $z=20$ as a function of $\alpha$, for different values of the cutoff parameter $\epsilon$, with $c_{\rm 200}$ set equal to 10. $x_{\rm ion}$ is almost independent of $\epsilon$ for small $\alpha$ since the luminosity is finite as $\epsilon \rightarrow 0$ for $\alpha < 1.5$. However, for larger values of $\alpha$, $x_{\rm ion}$ substantially increases as $\epsilon \rightarrow 0$. For the extreme case of $\alpha = 2, \epsilon = 10^{-5}$, the ionized fraction is nearly 30\%. Fig. \ref{fig1}(c) shows the dependance of $\delta\tau$ on the halo concentration parameter $c_{\rm 200}$, for $\alpha = 1.6,1.8,2$. The cutoff parameter $\epsilon$ was set  to $10^{-4}$. The solid red line indicates the maximum allowed value of $\delta\tau = 0.047$. The solid black line shows the expected Planck sensitivity $\tau_{\rm min} = 0.005$\cite{esa}. As expected, the model with $\alpha=2$ produces the most optical depth and is excluded for $c_{\rm 200} \gtrsim 7$. In all cases, the NFW model ($\alpha = 1$) can be expected to be negligible in influencing reionization and the optical depth. 

\subsection{Light dark matter candidates.}

We saw in the previous subsection that WIMPs with mass $\sim100$ GeV do not cause significant reionization, with the NFW density profile. Let us now consider lighter dark matter candidates, with masses in the MeV range. The cosmology of MeV dark matter has been considered by many authors\cite{h,h_mass,ak,beacom}. Studies of the soft gamma ray background\cite{ak} and the 511 keV line\cite{beacom} can be used to constrain the particle mass. Here, we investigate an independent approach, namely using the WMAP constraint on $\tau$ to place limits on the particle mass for a given cosmology. The form of $dN_\gamma/dx$ depends on the particle physics of the theory. In this article, we assume the same $dN_\gamma/dx$ as used previously for more massive WIMPs. Our results must therefore be treated with caution.

  We consider at first, halos with the NFW density profile, i.e. with $\alpha=1, \epsilon = 0$. The results are sensitive to the chosen minimum halo mass $M_{\rm min}$. $M_{\rm min}$ depends on the particle physics of the theory (in particular, the free streaming scale), and can be quite large ($\sim$ dwarf spheroidal galaxy) in certain models\cite{h_mass}. In those models, we do not expect to see a significant effect on reionization. However, since $M_{\rm min}$ is model dependent, we are justified in treating it as a free variable, and we consider models in which $M_{\rm min}$ varies from $10^{-6}$ to $1 \, M_\odot$. We also assume that $\langle \sigma_{\rm a} v \rangle = 3 \times 10^{-26}$ cm$^3$ s$^{-1}$ independent of $v$\cite{ak}. 
  
\begin{figure}[!ht]
\begin{center}
\scalebox{0.7}{\includegraphics{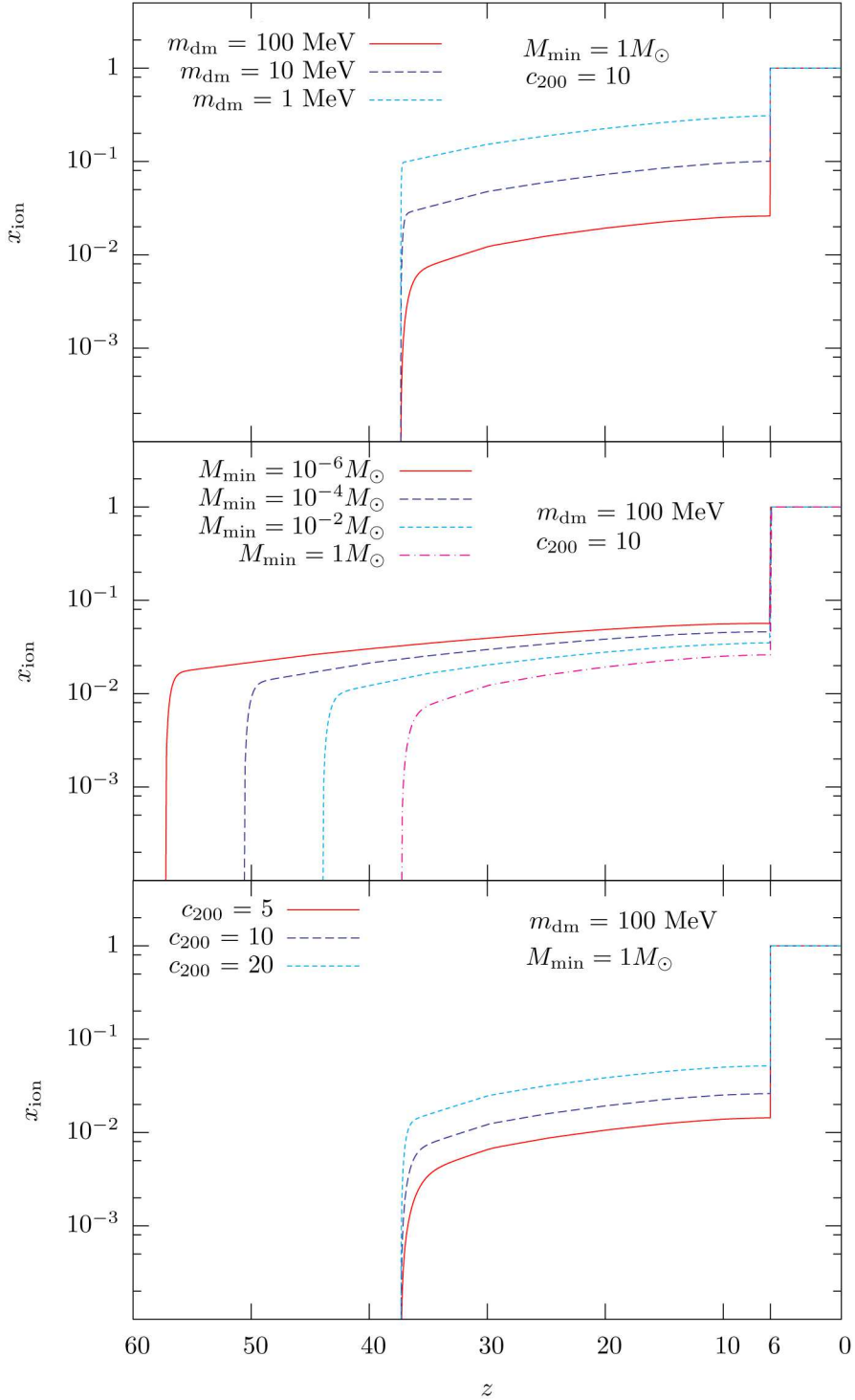}}
\end{center}
\caption{  (color online) Reionization due to halos composed of MeV scale dark matter particles. (a)  shows $x_{\rm ion}$ for different particle masses. $z_{\rm F}$ was set to 37.3 for $M_{\rm min} = 1\, M_\odot$. $x_{\rm ion} = 1$ for $z < 6$. Shown are results for particle masses $m_{\rm dm} = 100, 10,$ and 1 MeV. (b) shows the variation of $x_{\rm ion}$ with the minimum halo mass $M_{\rm min}$. (c) shows the dependence of $x_{\rm ion}$ on the concentration parameter.\label{fig2} }
\end{figure}

     Fig. \ref{fig2}(a) shows the ionized fraction $x_{\rm ion}$, for $m_{\rm dm} = 100, 10,$ and $1$ MeV, for an assumed minimum mass $M_{\rm min} = 1\,M_\odot$, and for $c_{\rm 200} = 10$. The halos are assumed to exist for $z > z_{\rm F}$, which in this case is $z_{\rm F}(1\,M_\odot) = 37.3$. The ionization calculation was carried out to $z=6$, beyond which, we set $x_{\rm ion} = 1$. We see that for MeV scale dark matter particles, the ionized fraction $x_{\rm ion}$ is significant (several percent) for NFW halos. There are two reasons that account for this increase. Firstly, a small particle mass results in a larger number density, and hence a higher probability of annihilation. Secondly, lighter dark matter particles produce photons with lower energy, resulting in a larger cross section for scattering. Fig. \ref{fig2}(b) shows the variation of $x_{\rm ion}$ with the minimum halo mass $M_{\rm min}$ (note that $z_{\rm F}$ depends on $M_{\rm min}$). Fig. \ref{fig2}(c) shows the dependence on $c_{\rm 200}$.

\begin{figure}[!ht]
\begin{center}
\scalebox{0.95}{\includegraphics{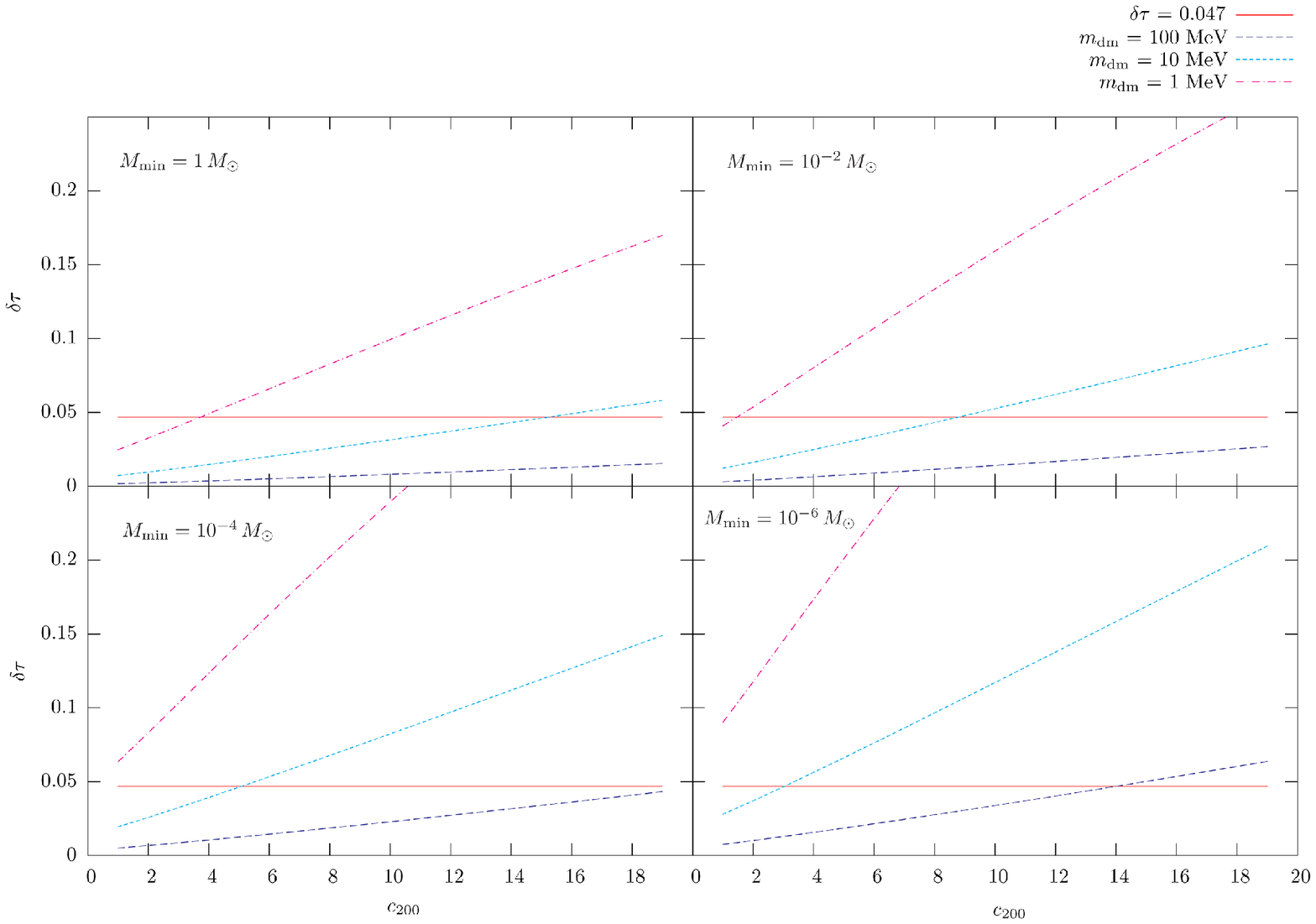}}
\end{center}
\caption{ (color online) Optical depth $\delta\tau$ as a function of the concentration parameter, considering halos made of MeV dark matter. Each panel shows a different value of the minimum halo mass $M_{\rm min}$.  The solid red line shows the maximum allowed value of $\delta\tau$. The solid black line indicates the expected sensitivity of the Planck experiment. The broken lines show the results for particle masses $m_{\rm dm} = 100,10,$ and 1 MeV. \label{fig3} }
\end{figure}

Let us now consider the contribution of dark matter halos to the optical depth, for MeV dark matter. Fig. \ref{fig3} shows $\delta\tau$ as a function of $c_{\rm 200}$ for different values of $m_{\rm dm}$ and $M_{\rm min}$. The four panels show the results for minimum halo masses $M_{\rm min} = 1, 10^{-2}, 10^{-4}$ and $10^{-6} M_\odot$. As expected, models with small $M_{\rm min}$ produce the most $\delta\tau$. The solid red line indicates the maximum allowed value of $\delta\tau = 0.047$, while the solid black line indicates the expected sensitivity of the Planck experiment. The broken lines show $\delta\tau$ for different particle masses $m_{\rm dm} = 1, 10$ and $100$ MeV. We see that very light ($\sim$ 1 MeV) dark matter candidates result in too much $\delta\tau$ for models with $M_{\rm min} < 10^{-2} M_\odot$, and are still disfavored for $M_{\rm min} = 1\,M_\odot$, except for very small values of $c_{\rm 200}$. We are thus able to constrain the particle mass for given halos and particle physics models.

\begin{figure}[!ht]
\begin{center}
\scalebox{0.75}{\includegraphics{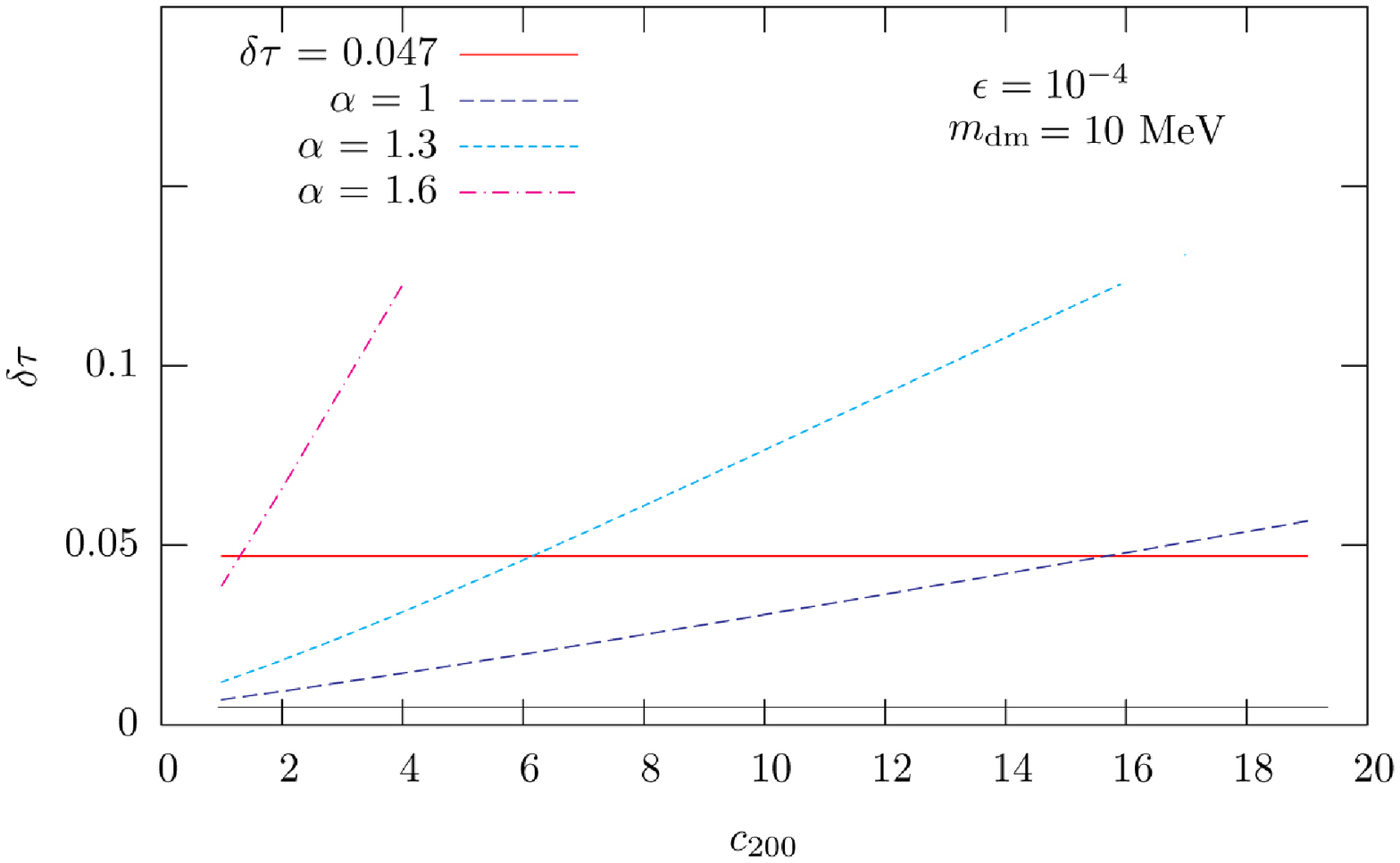}}
\end{center}
\caption{ (color online) Optical depth $\delta\tau$ as a function of the concentration parameter. Shown are cases $\alpha = 1, 1.3, 1.6$. $\alpha = 1$ corresponds to the NFW profile. The cutoff parameter $\epsilon = 10^{-4}, m_{\rm dm} = 10$ MeV, and $M_{\rm min} = 1 \, M_\odot$. The maximum allowed $\delta\tau = 0.047$ is indicated by the solid red line. The expected Planck sensitivity is shown by the solid black line.\label{fig4} }
\end{figure}

Consider now, the effect of varying the density profile power law exponent $\alpha$. Fig. \ref{fig4} shows $\delta\tau$ as a function of $c_{\rm 200}$ for a particle mass $m_{\rm dm} = 10$ MeV, and $M_{\rm min} = 1 \, M_\odot$. Three values of $\alpha$ are considered, namely $\alpha = 1, 1.3$ and $1.6$. The cutoff parameter $\epsilon$ was set equal to $10^{-4}$. For 10 MeV candidates, we see that large $\alpha \sim 1.6$ and small $\epsilon \sim 10^{-4}$ produce too much optical depth. As before, the solid red and black lines indicate the maximum $\delta\tau$ and the expected Planck sensitivity respectively.

\section{Conclusions.}

We fitted dark matter halos with a generic NFW profile parametrized by power law exponents $\alpha$ and $\beta$, concentration parameter $c_{\rm 200}$, and halo mass $M$. The luminosity of these halos was computed in terms of these parameters. $\langle \sigma_{\rm a} v \rangle = 3 \times 10^{-26}$ cm$^3$s$^{-1}$ was assumed to be independent of $v$. The Press-Schechter formalism was used to determine the number density of halos greater than a given minimum mass $M_{\rm min}$. We computed the ionization fraction as a function of redshift, produced by these dark matter halos. We then calculated the resulting optical depth and compared it to the WMAP measured value. 

 We first analyzed the effect of 100 GeV WIMPs (e.g. the lightest supersymmetric particle or the lightest Kaluza-Klein particle) on reionization. We found that if the dark matter halos follow an NFW profile, the effect on reionization and the optical depth is negligible. To have a considerable effect, the density profile needs to be considerably more cuspy ($\alpha \gtrsim 1.5$), with small cutoff $\epsilon \lesssim 10^{-4}$. 
 
 We then looked at lighter dark matter candidates, in the 1 to 100 MeV range which were found to be more promising. We found that NFW halos have a significant effect on the ionized fraction, with particle masses $m_{\rm dm} < 100$ MeV. We analyzed the contribution of dark matter halos to the optical depth, for different values of the minimum halo mass $M_{\rm min}$. For small $M_{\min} \sim 10^{-4} \, M_\odot$, we conclude that small particle masses $m_{\rm dm} < 10$ MeV are not favored. For larger $M_{\rm min} \sim 1\,M_\odot$, $m_{\rm dm} \sim 10$ MeV is allowed for small values of the concentration parameter $c_{\rm 200}$. The value of $M_{\rm min}$ is determined by the free streaming scale of the dark matter particles, which in turn depends on the scattering cross section of the dark matter particles with standard model particles in the early Universe. In order to determine $M_{\rm min}$, we would need to have a detailed understanding of the particle physics of the dark matter model. We can therefore use the predicted value of optical depth and future observations, to place constraints on MeV dark matter models. These limits provide information which may be used in conjunction with constraints on $m_{\rm dm}$ obtained by other means. 
 
   Our model of reionization with MeV dark matter in NFW halos suggests a uniform, and gradual reionization scenario, rather than a ``patchy'' or ``percolation'' scenario wherein ionized bubbles form first, and then merge together. Future observations of the 21 cm emission of neutral H may help distinguish this model from the patchy reionization model. A better understanding of the reionization process could also be relevant to dark matter substructure theories. A significant ionized fraction of gas could have important consequences for the formation of molecular H$_{\rm 2}$ which is the dominant coolant for primordial star formation. It would therefore be interesting to investigate whether early star formation could be influenced by a gradual reionization process such as the one we have described.  Our results are also relevant to theories where early star formation may be affected by dark matter annihilation\cite{stars}. Increased precision measurements of the CMB polarization would help us learn more about the reionization history. A detailed study of this epoch would lead to a better understanding of the first stars, and may also help us identify dark matter in the cosmic H-chamber.
    
\acknowledgments{We thank Jens Chluba, Dan Hooper, Richard Woodard, and Paolo Gondolo for helpful discussions. This work was supported by the Deutsche Forchungsgemeinschaft under grant GRK 881. }

\end{document}